\documentclass[conference]{IEEEtran}
\IEEEoverridecommandlockouts

\usepackage{amsmath,amssymb,amsfonts}
\usepackage{algorithmic}
\usepackage{graphicx}
\usepackage{textcomp}
\usepackage{xcolor}
\usepackage{pgfplots}
\usepackage{tabularx,array} 
\usepackage{cite}
\usepackage{hyperref}
\hypersetup{
    bookmarks=false,
    draft=true,
}
\def\BibTeX{{\rm B\kern-.05em{\sc i\kern-.025em b}\kern-.08em
    T\kern-.1667em\lower.7ex\hbox{E}\kern-.125emX}}

\pgfplotsset{
  ieee bar/.style={
    xbar,
    xmin=0, xmax=4.5,
    bar width=3.5pt,
    y=6.5mm,
    enlarge y limits={abs=4mm},
    axis x line*=bottom,
    axis y line*=left,
    xtick distance=1,
    y dir=reverse
  }
}

\definecolor{roleOEM}{RGB}{102,194,165}        
\definecolor{roleSupplier}{RGB}{252,141,98}   
\definecolor{roleTool}{RGB}{141,160,203}       
\definecolor{roleResearcher}{RGB}{231,138,195} 
\begin{document}
\begin{titlepage}
    \centering
    \vspace*{\fill}
    {\Large 
    This is the author’s version of the article that has been accepted for publication in IEEE ISSE 2026. The final version will be available at IEEE Xplore. 
    
    \vspace{1em}
    DOI: https://doi.org/10.1109/ISSE70179.2026.11763078.

\vspace{1em}
    
    © 2026 IEEE. Personal use of this material is permitted. Permission from IEEE must be obtained for all other uses, in any current or future media, including reprinting/republishing this material for advertising or promotional purposes, creating new collective works, for resale or redistribution to servers or lists, or reuse of any copyrighted component of this work in other works. 
}
    \vspace*{\fill}
\end{titlepage}
\title{Cross-Organizational SysML Model Integration: A Survey of Challenges and AI-Supported Tasks}

\author{\IEEEauthorblockN{1\textsuperscript{st} Li Zirui}
\IEEEauthorblockA{\textit{Product and Systems Engineering Group} \\
\textit{Technische Universität Ilmenau}\\
Ilmenau, Germany \\
https://orcid.org/0009-0007-7983-7901}
\and
\IEEEauthorblockN{2\textsuperscript{nd} Brix Torsten}
\IEEEauthorblockA{\textit{Product and Systems Engineering Group} \\
\textit{Technische Universität Ilmenau}\\
Ilmenau, Germany \\
torsten.brix@tu-ilmenau.de}
\and
\IEEEauthorblockN{3\textsuperscript{rd} Husung Stephan}
\IEEEauthorblockA{\textit{Product and Systems Engineering Group} \\
\textit{Technische Universität Ilmenau}\\
Ilmenau, Germany \\
https://orcid.org/0000-0003-0131-5664}
}

\maketitle

\begin{abstract}
Cross-organizational collaboration is widely regarded as a key promise of SysML-based Model-Based Systems Engineering (MBSE), yet practitioners still face persistent challenges when exchanging and integrating system models. In parallel, Large Language Models (LLMs) raise expectations for AI-assisted model understanding and integration, while reliability and required human oversight continue to pose challenges. This paper reports the results of an online questionnaire survey with 29 MBSE stakeholders involved in cross-organizational collaboration. Respondents rated eight predefined integration challenge categories and six AI-supported task types on five-point Likert scales. The results indicate that stakeholders perceive model integration as a multi-dimensional alignment problem across semantics, behavior, traceability, and exchange interoperability. These perceptions vary by organizational role and frequency of integration involvement. AI is rated highly useful for analysis tasks such as semantic structure analysis and inconsistency detection, and respondents predominantly prefer human-in-the-loop use with mandatory verification. These findings motivate AI support that enhances, rather than replaces, engineering responsibility in SysML-based integration.
\end{abstract}

\section{Introduction}
Cross-organizational collaboration in Model-Based Systems Engineering (MBSE) is becoming increasingly important, yet the exchange and integration of SysML-based models across company and tool boundaries remains a challenge due to heterogeneous modeling methods, tool ecosystems, and domain-specific practices among partners \cite{elaasar2025model,sohrt2025shared,prostep2023sysml}. 

In this context, System Modeling Language v2 (SysML v2) provides new technical capabilities for collaborative MBSE through formally defined semantics, semantic consistency between textual and graphical representations, and standardized REST API\cite{Bajaj.2022, Jansen.2022}. 
In parallel, recent work explores Large Language Models (LLMs)-based AI-assistance capabilities, such as model generation, model comprehension, and semantic conflict detection\cite{DeHart.2024}. However, the inherent uncertainty and limited explainability of LLM outputs pose challenges to their adoption in rigorous engineering practice\cite{mbse-ai.integration25,Topcu.2025}.

Empirical studies on the introduction of Systems Engineering in industrial contexts have identified persistent organizational, methodological, and tooling-related barriers that hinder effective adoption \cite{BRETZ2019783}. Furthermore, recent research on collaborative MBSE and model integration has made progress through survey-based investigations of MBSE stakeholders perspectives alongside literature-based and conceptual analyses\cite{Li.2024,elaasar2025model}. However, the perspectives of MBSE practitioners on cross-organizational model exchange and integration, as well as the role of emerging technologies such as AI, remain insufficiently explored \cite{henderson2021value}. Based on this motivation, the following research questions are formulated.

\begin{itemize}
\item RQ1: How do MBSE stakeholders perceive the severity of various challenge categories in the cross-organizational exchange and integration of SysML-based system models, and how do these perceptions relate to organizational roles and frequency of involvement in system model integration activities?

\item RQ2: How do MBSE stakeholders perceive the usefulness of AI-supported integration tasks, and which human-AI collaboration modes do they prefer?
\end{itemize}

To address the above research questions, this study designed and conducted a questionnaire survey targeting MBSE stakeholders, with a focus on model exchange and integration in cross-organizational collaboration scenarios. The survey collected responses from professionals representing different roles, including OEMs, suppliers, tool providers or consultants, and research institutions. Respondents provided their assessments using a five-point Likert scale and complemented their responses with open-ended questions to share practical experiences and expectations. 

The remainder of this paper reviews related work (Section~\ref{sec:stateofart}), describes the survey design and analysis (Section~\ref{sec:design}), reports the results (Section~\ref{sec:result}), and discusses implications and limitations (Section~\ref{sec:discussion}).

\section{State of the art}
\label{sec:stateofart}

\subsection{Model-Based Systems Engineering}
INCOSE Systems Engineering Vision 2035 highlights a shift toward model-centric digital engineering to address increasing system complexity, tight hardware–software integration\cite{SystemsEngineeringVision2035.6172025}. This shift positions models as lifecycle-wide information carriers that can support structured and traceable engineering information exchange across disciplines and organizational boundaries\cite{Miller.2022}.

In line with this vision, MBSE should support the early integration of models, facilitate cross-disciplinary collaboration, and improve development efficiency through model reuse\cite{Mahboob.2022b}. However, MBSE adoption in many organizations remains limited to local project or product levels, while enterprise-wide or cross-organizational integration is rarely achieved\cite{elaasar2025model}. 

From the perspective of MBSE system model languages, SysML v1 provides a unified means to represent requirements, structure, behavior, and constraints, and has been widely implemented in many modeling tools\cite{Friedenthal.2014}. As its application has expanded to multidisciplinary and cross-organizational collaboration scenarios, however, several limitations have become increasingly apparent.
On the one hand, SysML v1 is defined as a UML profile, which results in limited semantic precision. This has increased the difficulty of standardized usage and automated analysis\cite{Friedenthal.2023}. On the other hand, practical MBSE adoption is often characterized by organization- or project-specific methodologies and profile extensions\cite{estefan2007survey,elaasar2025model}. Differences among collaborating parties in abstraction levels further amplify the challenges of model alignment and reuse\cite{Li.2024}.

From an integration perspective, the profile-based nature of SysML v1 fosters variability in interpretation and tool-specific representation, which increases the likelihood of semantic drift when models are exchanged between organizations. This drift typically manifests as naming inconsistencies, mismatched abstraction levels, and diverging structural decompositions, thereby increasing manual reconciliation effort during integration. These recurring dimensions were used as a conceptual basis to structure the integration challenge categories assessed in the present survey.

Moreover, although model exchange mechanisms such as XMI formats have been established, existing surveys on model and data exchange indicate that tool-specific XMI implementations and the handling of additional metadata still require complex and context-dependent transformation and maintenance efforts when exchanging models across tools\cite{Zhou.2025}. 

Against this background, SysML v2 introduces a KerML-based semantic foundation that provides precise definitions of elements, relationships, and constraints\cite{Jansen.2022}. This supports improved consistency checking, model transformation, and formal analysis in comparison to profile-based approaches. This not only creates the basis for cross-tool transformation, configuration management, and cross-company collaboration in data-space-oriented scenarios, but also opens up new possibilities for integrating AI technologies into MBSE environments\cite{Li.2024,Li25LLM,CIBRIAN2025104350}.

\subsection{Cross-company collaboration scenario in MBSE}
In concrete cross-company collaboration scenarios, the OEM typically performs system decomposition and defines subsystem boundaries within its internal SysML models, resulting in black-box specifications that capture functional, structural, and interface aspects. Subsequently, subsystem-relevant model fragments are extracted from the global system model and delivered to suppliers using agreed exchange formats. Based on these specifications, suppliers carry out detailed design and implementation within their own methodologies and tool-chains, producing model-based representations of the subsystem that are then returned to the OEM for integration into the global system model, verification, and iterative refinement.\cite{prostep2023sysml,sohrt2025shared,Belkadi.2017}

Around this typical OEM–supplier workflow, both research and industrial practice have identified several classes of structural challenges. The use of a common modeling language alone is insufficient to ensure consistent understanding across organizational boundaries\cite{pandolf2023investigation}. Differences in abstraction levels, naming conventions among collaborating organizations can lead to divergent interpretations of the same model\cite{powley2020taking}. Furthermore, effective collaboration often requires explicit upfront agreements on modeling methods and interface specifications to achieve system integration \cite{benveniste2015contracts}. This OEM–supplier interaction highlights several structurally distinct integration layers such as model element alignment, structural \& behavioral consistency, and technical interoperability of exchange formats and tool infrastructures. Each of these layers may introduce independent sources of integration problems. 

\subsection{AI in MBSE}

In recent years, research has increasingly explored the use of AI technologies, particularly LLMs, to support MBSE activities\cite{SystemsEngineeringVision2035.6172025}. Proposed approaches include LLM-based assistants leveraging domain ontologies or the SysML metamodel to support requirements formulation, modeling guidance, and consistency checking, as well as the integration of LLMs with system models via the SysML v2 API to enable natural-language-driven model querying, modification, and the derivation of modeling elements\cite{mbse-ai.integration25,Ghanawi.2024,DeHart.2024}. 

In the context of model integration, AI applications can be differentiated into analysis support functions (e.g., model comprehension, semantic comparison, inconsistency detection), generative assistance (e.g., model element generation or transformation), and autonomous integration decisions\cite{li2025llm}. The acceptance of these categories may differ depending on accountability structures and contractual responsibility in cross-organizational engineering settings. This differentiation informed the selection and formulation of the six AI-supported task types evaluated in this study, covering primarily analysis and assistive functions as well as more generative forms of support.

While LLMs demonstrate capabilities in generating engineering artifacts, they also exhibit notable limitations, including data quality, semantic deviations, hallucinated content, and a lack of transparency\cite{hadi2023large,Hollender.2024}. These observations suggest that, within MBSE contexts, AI is more appropriately positioned as a support tool requiring review rather than as a fully autonomous agent\cite{Kulkarni.2024,Topcu.2025}.

\section{Survey Design and Data Analysis Method}
\label{sec:design}

In general, the state of the art highlights the technical diversity of MBSE approaches and modeling practices in cross-company collaboration scenarios, as well as the growing interest in AI-supported assistance. However, how these heterogeneous MBSE methods, tool ecosystems, and modeling practices and emerging capabilities are reflected in practical model exchange and integration remains unclear. To address this gap, an empirical questionnaire-based investigation was conducted to capture current practices and perceptions in SysML-based system model exchange and integration.

\subsection{Data collection approach and Survey Design}
The study is based on an online questionnaire to collect data. At the beginning of the survey, respondents were informed of the research purpose and anonymity, and the collected data were used for academic research and may be included in anonymized form in research publications.

The questionnaire was structured into three parts. The first part collected background information on respondents’ organizational roles, modeling languages and methodologies in use, and their level of involvement in cross-organizational model exchange and integration activities. The second and third parts addressed model-exchange challenges (RQ1) and AI-supported tasks and collaboration modes (RQ2), respectively. 

Survey items were derived from recurring challenge aspects and AI-related themes identified in the literature on MBSE collaboration, model integration, and AI-assisted Systems Engineering. Responses were collected using a combination of five-point Likert-scale items, multiple-choice questions, and a limited number of open-ended questions to capture additional practical insights.

The questionnaire was distributed via academic and industrial Systems Engineering communities primarily between May and July 2025 and yielded 30 responses, with 29 retained after response consistency and relevance checks. Detailed respondent backgrounds are summarized in Section~\ref{sec:background}. The original survey is available at\cite{GoogleFormsSurvey2026}.

With regard to validity, the questionnaire items were based on the OEM–supplier collaboration processes and model integration issues identified during the state of the art analysis presented in the section \ref{sec:stateofart}, which supports content validity. In addition, each category of challenges was accompanied by a brief illustrative example to help ensure a consistent understanding of the questions among respondents. Responses to the open-ended questions further complemented and corroborated the Likert-scale results by providing contextualized explanations and additional insights.

\subsection{Data Analysis Methods}

Generally, descriptive statistics (means, medians, and proportions of high ratings) were computed to characterize overall tendencies and perceived importance for Likert-scale items.
To explore differences in perceptions across respondent backgrounds, descriptive subgroup comparisons were conducted based on organizational role and frequency of involvement in model exchange and integration activities. Given the limited sample size and the ordinal data, these analyses focused on descriptive comparisons, examining differences in average ratings and distribution patterns across groups. The results are therefore interpreted in an exploratory manner and are not intended to support statistical inference or causal claims. 

Responses to open-ended questions were analyzed using a systematic coding procedure based on Mayring’s qualitative content analysis approach\cite{mayring2019qualitative}. An initial category framework was derived from the main model integration challenge aspects and AI application types identified in the literature review and applied to a subset of responses through pilot coding. Based on recurring and representative content, additional subcategories were inductively refined. The finalized coding scheme was then applied to all open-ended responses to identify practical challenges, application expectations, and risk concerns that the standardized questionnaire did not fully capture. All qualitative findings are reported in aggregated, anonymized form.

\section{Survey Result}
\label{sec:result}
This section presents the results of the questionnaire survey. First, it reports MBSE stakeholders’ assessments of the most prominent challenges encountered in SysML-based system model exchange and integration. Next, it presents findings on the perceived usefulness of AI-supported tasks, differences by engagement frequency and organizational role. Finally, it summarizes themes derived from open-ended responses.
\subsection{Respondent background and modeling context}
\label{sec:background}
The survey collected responses from 29 MBSE stakeholders with diverse organizational and functional backgrounds. In terms of organizational perspective, the sample includes 8 respondents with OEM-level system responsibility, 8 component or subsystem suppliers, 11 tool providers or consultants, and 6 academic researchers with several respondents reporting experience in more than one role.

Regarding functional responsibilities, respondents cover a broad range of roles along the Systems Engineering lifecycle, including system architects, requirements engineers, model developers, integration engineers, process or methodology specialists, as well as project managers and coordinators.

With respect to modeling languages, SysML v1.x is used by the large majority of respondents, while approximately half report experience with SysML v2. In addition, several respondents interact with UML, BPMN, ArchiMate, UAF, Capella, and company-specific modeling languages. A similar diversity is observed for modeling methodologies, including well-known approaches\cite{estefan2007survey} (e.g., MagicGrid, HarmonySE, OOSEM, ARCADIA, SPES) and company-specific variants.

Overall, the background data show that the survey captures practitioner perspectives from heterogeneous organizational settings characterized by mixed roles, multi-language tool-chains, and methodologically diverse MBSE practices, providing an appropriate context for the subsequent analysis of model integration challenges and attitudes toward AI-assisted integration.

\subsection{Challenges rating in model exchange and integration}

\begin{table}[htbp]
\caption{Model Integration Challenges}
\centering
\addtolength{\tabcolsep}{-1.5pt}
\small
\begin{tabularx}{\columnwidth}{|c|>{\raggedright\arraybackslash}X|c|c|c|}
\hline
\rotatebox{90}{\textbf{Label}}
& \rotatebox{90}{\textbf{Category}} 
& \rotatebox{90}{\textbf{Median}} 
& \rotatebox{90}{\textbf{4--5 Ratio}} 
& \rotatebox{90}{\textbf{Mean}} \\
\hline
A & Naming and Semantic Conflicts        & 4 & 65.5\% & 3.97 \\
B & Structural and Abstraction Challenges & 4 & 51.7\% & 3.55 \\
C & Interface and Port Integration Issues & 3 & 34.5\% & 3.24 \\
D & Behavioral and Logical Misalignment   & 4 & 58.6\% & 3.55 \\
E & Methodology and Profile Integration   & 3 & 35.7\% & 3.11 \\
F & Traceability Gaps                     & 4 & 58.6\% & 3.90 \\
G & Format Gaps (e.g., XMI vs JSON)       & 4 & 60.7\% & 3.68 \\
H & Incomplete/Incorrect Metadata         & 3 & 39.3\% & 3.25 \\
\hline
\end{tabularx}
\label{challenges}
\end{table}

Table \ref{challenges} summarizes respondents’ ratings of eight categories of challenges (A–H) encountered during cross-organizational SysML model exchange and integration. All items were rated using a five-point Likert scale (1 = Not an issue, 5 = Very significant issue). For each challenge, the median, the proportion of high ratings (scores of 4 or 5), and the mean value are reported.

Naming and semantic conflicts (A), behavioral and logical inconsistencies (D), missing traceability relationships (F), and differences in model formats (G) consistently receive high ratings, with medians of 4. For these aspects, roughly 60\% of respondents assign scores of 4 or 5, with mean values between 3.55 and 3.97, indicating that they are widely perceived as prominent integration issues.
Structural and abstraction challenges (B) also show a median of 4, but with a lower proportion of high rating (51.7\%) and a mean of 3.55. 

In contrast, interface and port integration issues (C), methodological and SysML profile integration (E) and metadata-related issues (H) show medians of 3. For these aspects, the proportion of high ratings ranges from 34.5\% to 39.3\%, with mean values approximately between 3.11 and 3.25, reflecting a more dispersed distribution of responses.

Overall, the results suggest that challenges in cross-organizational SysML system model integration are not confined to a single technical aspect, but span multiple aspects including semantics, consistency, formats, and methodologies. However, the degree of consensus among respondents varies substantially across these challenges.

\subsection{AI support usefulness and oversight preferences}

\begin{table}[htbp]
\caption{AI-Supported Model Understanding and Integration Tasks}
\centering
\addtolength{\tabcolsep}{-2.5pt}
\small
\begin{tabularx}{\columnwidth}{|c|>{\raggedright\arraybackslash}X|c|c|c|}
\hline
\textbf{Label} 
& \textbf{Category} 
& \textbf{Median} 
& \textbf{4--5 Ratio} 
& \textbf{Mean} \\
\hline
A & Understand model content from textual or structural input 
  & 5 & 75.9\% & 4.28 \\
B & Suggest mappings between model elements 
  & 4 & 79.3\% & 4.21 \\
C & Identify semantic conflicts or inconsistencies 
  & 5 & 82.8\% & 4.34 \\
D & Generate integration workflows and prompts 
  & 4 & 65.5\% & 3.72 \\
E & Suggest cross-methodology transformations 
  & 4 & 55.2\% & 3.55 \\
F & Support semantic structure analysis 
  & 5 & 89.7\% & 4.45 \\
\hline
\end{tabularx}
\label{ai_supported_tasks}
\end{table}

Table \ref{ai_supported_tasks} summarizes respondents’ assessments of the potential usefulness of AI across six model exchange and integration–related tasks derived from challenges in last section. All tasks were assessed using a five-point Likert scale (1 = Not useful, 5 = Very useful), and the results are reported in terms of median values, the proportion of high ratings (scores of 4 or 5), and mean scores.

Tasks related to model understanding and semantic analysis received the most consistently high ratings. In particular, semantic structure analysis support (F) and identification of semantic conflicts or inconsistencies (C) both show a median rating of 5, with high-rating proportions above 80\%.
Similarly, tasks associated with model content comprehension (A) and model element mapping suggestions (B) were also rated highly. Both tasks exhibit high-rating proportions above 75\%, with mean scores of 4.28 and 4.21.
By contrast, generation of integration workflows and guidance (D) and cross-methodology transformation suggestions (E) are still viewed positively but receive lower overall ratings, with medians of 4 and noticeably lower proportions of high ratings.

Overall, the results indicate a strong consensus on the potential value of AI for analysis, interpretive, and conflict-detection tasks. However, generative tasks, such as creating integration workflows or cross-methodology transformations are perceived as less beneficial overall, though the assessments remain positive.

Regarding acceptable levels of human oversight in AI-assisted tasks, majority of respondents (83\%) preferred a semi-automatic mode in which AI provides support with mandatory human verification. A further 17\% favored a more conservative setup in which AI is limited to making suggestions, with the primary work remaining manual.

Notably, no respondents selected either of the two extremes: manual workflows without AI support or highly automated modes with optimal review. This indicates a clear preference for human-led integration with AI in an assistive role.

\subsection{Engagement-Dependent Differences in Challenge Perception}
\begin{figure}[!t]
  \centering
  \begin{tikzpicture}

    \begin{axis}[
        width=0.9\columnwidth,
        height=4.6cm,
        xbar,
        xmin=0,xmax=4.5,
        xlabel={Average Rating},
          xlabel style={
    at={(axis description cs:0.5,0.02)},
    anchor=north
  },
        symbolic y coords={A,B,C,D,E,F,G,H},
        ytick=data,
        y dir=reverse,
        yticklabels={
          {A. Naming},
          {B. Structure},
          {C. Interface},
          {D. Behavioral},
          {E. Method},
          {F. Traceability},
          {G. Format},
          {H. Metadata}
        },
        yticklabel style={font=\scriptsize},
        bar width=3pt,
        enlarge y limits=0.1,
        axis x line*=bottom,
        axis y line*=left,
        xtick distance=1,
        legend style={
          at={(-0.25,1.1)},
          anchor=west,
          legend columns=1,
          font=\scriptsize,
          draw=none,
          fill=none
        }
      ]

      \addplot+[fill=teal!70,draw=teal,bar shift=+1pt] coordinates {
        (3.857142857,A)
        (3.571428571,B)
        (3.071428571,C)
        (3.785714286,D)
        (3.357142857,E)
        (3.785714286,F)
        (3.92307692,G)
        (3.76923077,H)
      };
      \addplot+[fill=orange!85,draw=orange,bar shift=-2.7pt] coordinates {
        (4.066666667,A)
        (3.533333333,B)
        (3.4,C)
        (3.333333333,D)
        (2.857142857,E)
        (4.0,F)
        (3.46666667,G)
        (2.8,H)
      };

      \legend{High (Frequently/Occasionally), Low (Rarely/Never)}
    \end{axis}

  \end{tikzpicture}
  \vspace{-6pt}
  \caption{Perceived challenge severity by cross-organizational engagement level}
  \label{fig:engagement}
\end{figure}
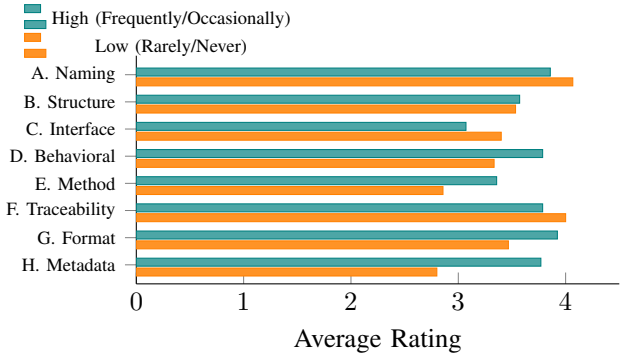

Figure \ref{fig:engagement} compares respondents’ ratings of the eight integration challenges under different frequency of engagement in model exchange and integration activities. In this survey, "engagement" refers to the self-reported frequency of involvement in cross-organizational model exchange and integration. 14 respondents who selected "frequently" or "occasionally" are grouped as high engagement, while 15 respondents who selected “rarely” or “never” are grouped as low engagement. Ratings are reported as group-level mean values.

Across challenges, respondents with higher involvement tend to assign higher importance to behavioral misalignment (D), format gaps (G) and metadata issues (H). Differences in perceived difficulty for structural abstraction (B) and methodology or profile integration (E) remain small between high- and low-engagement groups. For naming and semantic conflicts (A), Interface and Port Integration Issues (C) and traceability gaps (F), ratings are slightly higher in the low-engagement groups.

Overall, differences between engagement-frequency groups are more evident for challenges related to model dynamic consistency and exchange artifacts. Meanwhile, ratings for most basic modeling challenges remain relatively similar across groups.

\subsection{Role-Specific Differences in Challenge Perception}

\begin{figure*}[!t]
  \centering
  \begin{tikzpicture}
    \begin{axis}[
        ieee bar,  
        yticklabel style={font=\scriptsize},
        width=0.9\textwidth,
        height=6cm,
  xlabel={Average Rating},
  xlabel style={
    at={(axis description cs:0.5,0.02)},
    anchor=north
  },
        symbolic y coords={A,B,C,D,E,F,G,H},
        ytick=data,
        yticklabels={
          {A. Naming/Semantic},
          {B. Structure/Abstraction},
          {C. Interface/Ports},
          {D. Behavioral},
          {E. Method/Profiles},
          {F. Traceability},
          {G. Format Gaps},
          {H. Metadata}
        },
        legend columns=4,
        legend style={at={(0.5,1)}, anchor=south, font=\scriptsize, draw=none, fill=none}
      ]

      \addplot[fill=roleOEM, draw=none, bar shift=+6pt] coordinates{
        (4.25,A)(3.375,B)(2.875,C)(3.5,D)(2.857142857,E)(3.875,F)(3.25,G)(2.625,H)
      };
      \addplot[fill=roleSupplier, draw=none, bar shift=+2pt] coordinates{
        (3.875,A)(4.0,B)(3.625,C)(3.75,D)(3.75,E)(3.75,F)(3.857142857,G)(3.571428571,H)
      };
      \addplot[fill=roleTool, draw=none, bar shift=-2pt] coordinates{
        (3.818181818,A)(3.363636364,B)(3.0,C)(3.363636364,D)(2.909090909,E)(4.181818182,F)(3.909090909,G)(3.363636364,H)
      };
      \addplot[fill=roleResearcher, draw=none, bar shift=-6pt] coordinates{
        (3.833333333,A)(3.333333333,B)(2.833333333,C)(3.666666667,D)(3.0,E)(3.833333333,F)(3.333333333,G)(3.333333333,H)
      };

      \legend{OEM, Supplier, Tool Vendor, Researcher} 

    \end{axis}
  \end{tikzpicture}
  \vspace{-6pt}
  \caption{Perceived challenge severity across respondent roles}
  \label{fig:challenge-role}
\end{figure*}
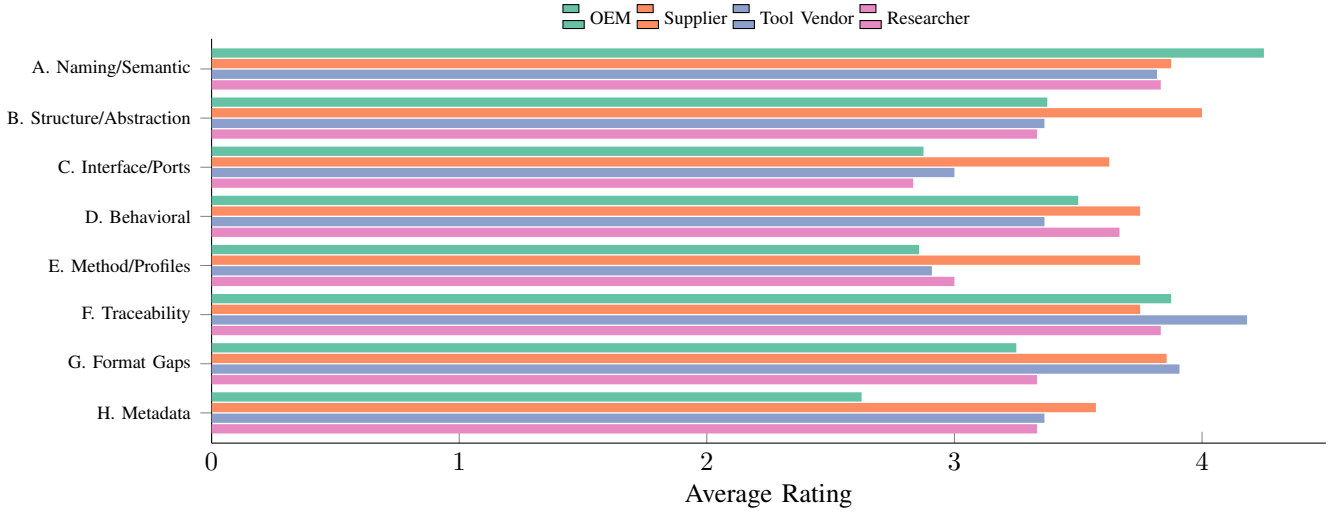

Figure \ref{fig:challenge-role} compares how different organizational roles (OEMs, suppliers, tool vendors, and researchers) perceive the challenges A–H in collaborative SysML-based system model exchange and integration. Since the organizational role was collected as a multi-select item, some respondents reported experience in more than one role (e.g., OEM and researcher). As a result, the role-based groups partially overlap. The results should therefore be interpreted as ratings of respondents with experience in a given role, rather than as comparisons between mutually exclusive groups. In the final sample, 8 respondents reported OEM, 8 suppliers, 11 tool vendors or consultants, and 6 researchers. The results indicate role-related differences in the perceived importance of several challenge aspects (A–H).

Across all roles, Challenge A receives consistently high ratings, with OEMs assigning relatively higher importance than the other groups, whose evaluations remain closely aligned. In contrast, clearer role-specific results emerge for B and C, where suppliers report higher perceived difficulty than OEMs and researchers, while tool vendors occupy an intermediate position. For D, the ratings are largely comparable across roles.

More pronounced divergence is observed for E, where suppliers again report higher concern than OEMs and tool vendors, with researchers positioned between these groups. A distinct pattern appears for F, which is rated higher by tool vendors than by the other roles. Finally, for G and H, suppliers and tool vendors consistently assign higher importance than OEMs, with researchers again showing intermediate evaluations.

Overall, Figure \ref{fig:challenge-role} shows that, while the general rating trends remain similar across all roles, differences between organizational roles primarily manifest at the level of individual challenge aspects (A–H).

\subsection{Themes from open-ended responses}

The open-ended questions were analyzed using a qualitative content analysis approach based on Mayring. The analysis resulted in a set of themes that complement the predefined challenge aspects and AI-related items covered by the closed questions. For clarity, the results are presented according to the three open-ended questions in the questionnaire.

\noindent\textbf{Additional Challenges in Model Exchange and Integration}

Respondents identified several additional challenges that extend beyond the predefined integration aspects. A prominent aspect concerns tool limitations and interoperability, including restricted access to model data, unstable exchange mechanisms, and practical difficulties with standardized formats.

A second theme relates to configuration, change, and consistency management, encompassing issues such as tracking changes across model versions, maintaining consistency between models and related lifecycle artifacts, and handling model evolution across iterations.

Respondents also emphasized process-, organizational-, and governance-related constraints, including late identification of the need for model exchange, insufficient integration of model exchange into information management strategies, and legal or contractual requirements that finally enforce document-based deliverables.

Finally, several responses pointed to granularity and perspective mismatches, where differences in abstraction levels or disciplinary viewpoints complicate integration, even when shared modeling tools are used.

\noindent\textbf{Promising AI Use Cases for Model Integration}

Regarding AI support, respondents highlighted several classes of promising Use Cases. One recurring theme describes AI-assisted model understanding and analysis, including AI-based copilot functionality, natural-language querying of models, and support for detecting semantic conflicts or inconsistencies.

Another frequently mentioned theme focuses on the automation of repetitive and low-level modeling tasks, such as correcting notation errors, reducing manual interaction effort, and supporting rapid model validity checks in large models.

Respondents also pointed to model generation and transformation support as a promising area, including the transformation of informal or textual inputs into SysML models, automated diagram creation or refactoring, and transformations across methodologies or SysML versions.

A further theme concerns traceability and cross-domain integration, where AI is seen as supporting end-to-end traceability, consistency analysis, and impact assessment across models, requirements, and related lifecycle artifacts.

\noindent\textbf{ Additional Comments and Suggestions on AI Assistance}

In the final open-ended question, respondents frequently addressed preconditions and risks for AI adoption, emphasizing that AI should not be used to compensate for poor data quality or insufficient modeling discipline, and that fully automated solutions should be approached with caution.

Another theme highlights the role of AI in knowledge structuring and representation, including transforming unstructured information into formal representations and enabling more intuitive access to structured knowledge through natural-language interaction.

Finally, respondents stressed tool-chain integration and architectural requirements for effective AI assistance, such as the need for standardized API, write-back capabilities into engineering tools, and closer integration of SysML-based MBSE tools with related environments such as ALM, PLM, and Digital Twin platforms.

\section{Discussion}
\label{sec:discussion}
This section interprets the survey findings in relation to existing MBSE and AI-assisted Systems Engineering research. While the preceding sections reported the empirical results of the questionnaire, the following discussion relates these observations to prior work and interprets them in the context of the state of the art. It focuses on explaining why certain integration challenges are perceived as particularly prominent, how these perceptions vary across roles and experience levels, and what the findings imply for the design of AI-assisted model integration in SysML-based MBSE contexts.

\subsection{Core integration challenges in SysML-based MBSE}
The survey results indicate that SysML-based model integration is experienced as a multi-dimensional alignment challenge spanning semantics, behavior, traceability, and exchange formats, rather than a single technical bottleneck. The consistently high ratings of these aspects suggest that integration challenges emerge across multiple layers of the system model simultaneously.

In the responses, issues related to naming and semantics remain particularly prominent. From an interpretative perspective, inconsistent terminology and discipline-specific interpretations may lead to divergent understandings of ostensibly shared model elements, even when SysML is adopted on both sides of OEM and supplier. Structural and abstraction-level mismatches may further compound this problem, as supplier subsystem models often reflect implementation-oriented viewpoints that are difficult to reconcile with OEM-level architectural decompositions. Behavioral misalignment can introduce complexity, especially in iterative integration scenarios, where asynchronous model evolution may result in inconsistencies between expected and realized system behavior. Traceability gaps and format-related issues suggest that integration challenges extend beyond model content to encompass lifecycle information management and exchange infrastructure.

In this context, the more formal semantic foundation of SysML v2 provides an important basis for reducing ambiguity and supporting automated analysis and consistency checking\cite{Jansen.2022}. However, the results indicate that formalized language semantics alone are not enough for cross-organizational collaboration. Explicit alignment mechanisms may help to ensure semantic clarity throughout the collaboration process.

Open-ended responses on format gaps and metadata issues highlight concerns about fragile XMI exchanges and tool-specific transformations, which underline the need for more robust exchange approaches. Standardized API, as proposed in SysML v2, may help to enable controlled access to model content and support continuous integration across organizational boundaries, shifting collaboration toward more sustainable integration workflows.

Finally, the AI-related findings suggest that AI can support cross-organizational integration primarily in an assistive role. MBSE stakeholders see clear value in AI for model understanding, semantic analysis, and conflict detection, while expressing caution toward highly generative or autonomous transformations. This indicates that AI can be positioned as a complementary layer that enhances transparency and reduces cognitive load, embedded within well-defined semantic, organizational, and technical integration frameworks.

\subsection{Role- and experience-specific perspectives}
The analysis of subgroup differences reveals that perceived integration challenges vary with both MBSE stakeholders’ level of engagement in integration activities and their organizational roles, indicating that integration difficulties are shaped by practical exposure and responsibility within the collaboration workflow.

With respect to engagement frequency, respondents with lower involvement in cross-organizational tasks tend to rate naming and semantic conflicts as well as traceability issues slightly higher than those with frequent engagement. In contrast, highly engaged respondents consistently assign higher importance to challenges related to behavioral alignment, model format, methodological and profile integration, and metadata consistency. Meanwhile, differences related to structural abstraction and interface integration remain comparatively small across engagement groups. One possible explanation is that occasional integration efforts are particularly exposed to terminology and traceability issues during initial alignment, while frequent integration activities bring cumulative experience with behavioral, format, and metadata-related problems to the foreground.

Role-based comparisons indicate that perceived importance of the different challenge aspects varies with organizational responsibilities. OEM respondents assign the highest importance to naming and semantic conflicts, consistent with their system-level responsibility for consolidating models from multiple suppliers and disciplines. Suppliers exhibit a more evenly distributed concern profile, but report particularly high difficulty with structural and abstraction mismatches as well as methodology and profile integration, reflecting the effort required to align implementation-level models with OEM-defined system architectures and methods. Tool providers and consultants are particularly sensitive to traceability gaps and format-related issues, in line with their role in supporting interoperability across heterogeneous tools and maintaining data continuity during model exchange. Researchers display comparatively balanced ratings across aspects, with slightly elevated attention to semantic, behavioral, and traceability issues, which may reflect exploratory modeling contexts and diverse methodological research.
Overall, these results suggest that integration support is likely to be more effective when it is role-aware rather than generic—for instance, by prioritizing semantic harmonization for OEMs, methodology alignment for suppliers, and robust traceability and exchange infrastructures for tool providers and consultants.

\subsection{Insights for AI-assisted integration}
The survey results indicate a broadly positive but differentiated attitude toward the use of AI in SysML-based model exchange and integration, with stakeholders assigning higher value to analysis and supportive tasks such as model comprehension, semantic structure analysis, and inconsistency detection. These results suggest that AI is primarily perceived as a means to enhance engineers’ ability to understand and assess complex, distributed models rather than to replace established integration practices.

In contrast, AI use cases that involve automated generation of integration workflows or suggestions for cross-methodology alignment receive lower and more heterogeneous ratings. This may reflect a cautious stance toward AI interventions that directly influence modeling decisions or methodological alignment, particularly in cross-organizational contexts where responsibility and contractual boundaries are clearly defined. 

Preferences regarding human–AI collaboration further clarify this distinction. Most respondents favor human-leading interaction modes, in which AI provides recommendations or partial automation but all outcomes remain subject to mandatory human review. Notably, neither fully manual approaches that exclude AI support nor highly automated modes with minimal human oversight gain support, indicating that MBSE stakeholders seek a balanced integration of AI and human decision while preserving control and responsibility.

\subsection{Insights from Open-Ended Responses}
The open-ended responses suggest that cross-organizational SysML-based system model integration requires coordinated advances in semantics, infrastructure, and tooling. While SysML v2 offers a stronger formal semantic basis, effective collaboration additionally depends on explicit alignment mechanisms.

Persistent concerns regarding format gaps and fragile exchanges highlight the need for API-centric, tool-independent integration infrastructures that replace file-based model transfer and enable controlled access to model data. At the same time, the importance of traceability and consistency challenges indicates that integration support must include lifecycle artifacts and version management.

Finally, the AI-related results point to a clear design boundary: AI is most acceptable as an analysis assistant supporting model understanding, semantic analysis, and traceability, while structurally invasive or generative decisions should remain under human control.

\subsection{Limitations and future work}
This study has several limitations that should be considered when interpreting the results. First, the sample size is relatively small and based on voluntary participation within specific MBSE communities, which may introduce selection bias and limits the extent to which the findings can be generalized. Second, the findings rely on self-reported perceptions measured on ordinal Likert scales, and thus reflect subjective experience rather than objective integration performance. Third, role-based analyses are affected by overlapping role assignments and uneven group sizes, which constrains the strength of comparative conclusions.

Future work will focus on developing and evaluating concrete model alignment mechanisms for cross-organizational SysML-based integration, addressing semantic, behavioral, and traceability consistency across heterogeneous tools and abstraction levels. In this context, AI-assisted approaches will be explored as supportive mechanisms—for example, to assist semantic alignment, detect inconsistencies, and maintain traceability—while keeping integration decisions under explicit human control.

\section{Conclusion}
\label{sec:conclusion}
This paper provides empirical insights into how MBSE stakeholders perceive integration challenges and AI opportunities in SysML-based system model exchange across organizational boundaries. Addressing RQ1, the findings show that integration difficulties arise from semantic, behavioral, traceability, and format-related issues, whose relative importance varies with organizational role and integration experience.

Regarding the usefulness of AI-supported tasks (RQ2), stakeholders attribute high value to AI support for analysis tasks such as model understanding, semantic conflict detection, and traceability analysis, while expressing more cautious attitudes toward generative transformations and highly automated workflows. Open-text responses further highlight design requirements for future integration solutions, including robust standards and API, semantically enriched data infrastructures, and close integration with lifecycle management and Digital Twin environments.

Overall, the paper contributes practitioner-grounded insights to ongoing discussions on SysML v2–based collaboration and AI-assisted Systems Engineering. It supports a view of AI as an embedded, assistive component in MBSE tool-chains that enhances transparency and reduces routine workload under human oversight. The results point toward concrete research directions for building trustworthy, role-sensitive integration support in future MBSE ecosystems.

\bibliographystyle{IEEEtran}
\bibliography{reference.bib}
\end{document}